\documentclass[12pt]{article}
\usepackage{epsf}
\title{\bf $\rm\bf QCD_2$ in the modified Fock--Schwinger gauge}
\author{Yu.S.Kalashnikova\thanks{e-mail: yulia@vxitep.itep.ru},
A.V.Nefediev\thanks{e-mail: nefediev@vxitep.itep.ru}}
\date{\it Institute of Theoretical and Experimental Physics, 117218, Moscow,
Russia}
\newcommand{\ds}{\displaystyle}
\newcommand{\be}{\begin{equation}}
\newcommand{\ee}{\end{equation}}
\newcommand{\vpint}{\int\makebox[0mm][r]{\bf --\hspace*{0.13cm}}}
\newcommand{\too}{\mathop{\to}\limits_{N_C\to\infty}}
\newcommand{\trr}{\mathop{Tr}\limits_{x\to y+}}
\begin{document}
\maketitle

\begin{abstract}
The system of light quark and heavy antiquark source is studied in 1+1 QCD
in the large $N_C$ limit. The situation is demonstrated to be
the two dimensional analogue of the problem in 3+1 QCD corresponding to the 
essentially nonlocal case of zero vacuum correlation length, which does not
allow perturbative treatment.
The modified Fock--Schwinger gauge condition is employed to derive the 
effective Dirac-type equation for the spectrum, 
which is proved to be equivalent to the 
't~Hooft equation in the limit of one heavy mass. The chiral condensate is 
shown to appear naturally in the given approach and to coincide with the 
standard value. 
\end{abstract}

Theory of quarks in 1+1 dimensions interacting via gluons in the
large $N_C$ limit has a long history. It was in 70-th when the
celebrated 't Hooft equation for the meson spectrum in such a model
was derived in the light-cone gauge ($A_-=0$) \cite{1} and then 
re-derived in the axial gauge ($A_1=0$) \cite{2}. Since that time
it has done it credit to every new approach to the realistic 3+1
QCD to reproduce this equation when applied to a much more simple
and physically clear 1+1 't Hooft model. Many theories
have passed this examination with honour.

Recently there was suggested a new approach to the theory of 
light spinning quarks in the confining vacuum \cite{3}, that 
allowed, in the large $N_C$ limit, to 
obtain a nonlinear Dirac-type equation for the light quark motion 
in the
nontrivial vacuum of QCD in presence of heavy antiquark source.
Despite of its nonlocality, this
equation allows a systematic analysis, which was started in our
previous work \cite{4} and led to the conclusion that the 
two cases of small and large vacuum correlation length $T_g$, 
value which defines the distances where vacuum gluonic fields are
correlated, differ drastically, {\it viz.} for the case of large 
$T_g$ ($T_g\gg \frac{1}{m}$,
where $m$ is the mass of the light quark) the nonlocality can be
treated perturbatively, whereas the opposite limiting case of small
$T_g$ ($T_g\ll \frac{1}{m}$) turned out sufficiently nonlocal and
demanded extra ideas for its considering. Fortunately the 't Hooft
model is of great help at this point. 

Indeed, in 3+1 the interaction of the
light quark with the heavy antiquark is driven by the irreducible 
correlators $<A_{\mu_1}^{a_1}(x_1)\ldots
A_{\mu_n}^{a_n}(x_n)>$, where $a_1,\ldots,a_n$ are the colour indices 
in the adjoint representation.
No exact solutions derived directly from the QCD are known for such 
correlators, and the strategy \cite{3} is to employ a kind of 
the Fock--Schwinger gauge which allows to express correlator 
$<A\ldots A>$ in terms of the gauge--invariant field strength one,
$<F\ldots F>$, and to insert some theoretical input for the latter.
Namely, it was shown in the framework of the Vacuum Background Correlators 
method \cite{6} that with the Euclidean bilocal field strength correlator 
\be
<F^a_{\mu\nu}(x)F^b_{\lambda\rho}(y)>=\delta^{ab}\frac{2N_C}{N_C^2-1}D(x-y)
(\delta_{\mu\lambda}\delta_{\nu\rho}-\delta_{\mu\rho}\delta_{\nu\lambda})
\label{5}
\ee
normalized by the condition
\be
\sigma=2
\int_0^\infty d\tau\int_0^\infty d\lambda D(\tau,\lambda)
\label{6}
\ee
the area law asymptotic for the isolated Wilson loop average holds true,
$<W(C)>\to N_C{\it exp}(-\sigma S)$. Expression (\ref{6}) for the string 
tension implies that function $D$ decreases rapidly in all directions
of the Euclidean space, and this decrease is governed by gluonic 
correlation length $T_g$.

We are interested in the limit $T_g\to 0$. It is easy to see that quite 
the same situation is realised in 1+1 QCD, where the Coulomb 
force is confining and in any axial gauge the only nontrivial correlator 
of gluonic 
fields is the gluon propagator which is bilocal and
effectively corresponds to the case of $T_g$ equal to zero. 

Indeed, the bilocal field strength correlator in 1+1 case,
$<0|TF_{10}^a(x_0,x)F_{10}^b(y_0,y)|0>$, is gauge--invariant and can be 
calculated for example in the axial gauge $A_1^a=0$ with the result
\footnote{We have included the coupling constant $g$ into the definition
of the correlator.}
\be
i<0|TF_{10}^a(x_0,x)F_{10}^b(y_0,y)|0>=-g^2\delta^{ab}\delta^{(2)}(x-y).
\label{210}
\ee

Then there exists the relation in the Euclidean space, which is similar to 
(\ref{5}):
\be
<F^a_{10}(x)F^b_{10}(y)>=\delta^{ab}\frac{2N_C}{N_C^2-1}{\tilde D}(x-y),
\label{220}
\ee
where 
\be
{\tilde D}(x-y)=\frac{N_C^2-1}{2N_C}g^2\delta^{(2)}(x-y)=\lim_{T_g\to 0}
D\left(\frac{x_0-y_0}{T_g},\frac{x-y}{T_g}\right)
\ee
is normalized by the condition
\be
\sigma=2
\int_0^\infty d\tau\int_0^\infty d\lambda {\tilde D}(\tau,\lambda)=
\frac{g^2}{4}\left(N_C-\frac{1}{N_C}\right)\too \frac{g^2N_C}{4}
\ee

The latter observation allows us to kill two birds with one stone, 
when applying
the given approach to the two-dimensional QCD. First, we prove the
selfconsistency of the general method under consideration. On the
other hand, dealing with nonlocal interaction in 1+1 we gain
valuable experience which may be extremely useful for solving the 
problem for the physical four dimensional case in the most
interesting limit of small $T_g$. The QCD string is believed to
be formed in this limit, and we plan to highlight this question in the
future publications.

As in the 3+1 case \cite{3,4} we start from the Green function 
$S_{q\bar{Q}}$ for the $q\bar{Q}$ system in Minkowski space-time
\footnote{We use the following $\gamma$-matrix convention:
$\gamma_0=\sigma_3$, $\gamma_1=i\sigma_2$, $\gamma_5=\sigma_1$.}
\be
S_{q\bar Q}(x,y)=\frac{1}{N_C}\int
D{\psi}D{\psi^+}DA_{\mu}
\exp{\left\{-\frac14\int 
d^2x
F_{\mu\nu}^{a2} 
-\int
d^2x
{\psi^+}(i\hat \partial -m -\hat A)\psi \right\}}\times
\label{1}
\ee
$$
\times\psi^+(x) S_{\bar Q} (x,y)\psi(y),
$$
where we have placed the static antiquark at the origin, and
$S_{\bar Q} (x,y)$ denotes its propagator. In the one--body limit 
we can considerably 
simplify the situation making use of the modified Fock--Schwinger
gauge \cite{6}, having set
\footnote{In 3+1 this gauge is usually introduced via conditions
$A^a_0(x_0,\vec{0})=0$ and $\vec{x}\vec{A}^a(x_0,\vec{x})=0$,
which obviously reduce to (\ref{2}) in 1+1.}
\be
A^a_0(x_0,0)=0,\quad A^a_1(x_0,x)=0.
\label{2}
\ee

In such a gauge the gluon field is expressed in terms of the field 
strength tensor as
\be
A_0^a(x_0,x)=\int_0^1 d\alpha xF^a_{10}(x_0,\alpha x),\quad A^a_1(x_0,x)=0,
\ee
and the gluon propagator takes the form
\be
K^{ab}_{11}(x_0-y_0,x,y)=K^{ab}_{01}(x_0-y_0,x,y)=0
\ee
$$
K^{ab}_{00}(x_0-y_0,x,y)=xy\int_0^1d\alpha\int_0^1 d\beta 
<0|TF^a_{10}(x_0,\alpha x)F^a_{10}(y_0,\beta y)|0>=
$$
\be
=-\delta^{ab}\frac{g^2}{2}\delta(x_0-y_0)(|x|+|y|-|x-y|)\equiv 
\delta^{ab}K(x,y)
\label{3}
\ee

In what follows we use expression (\ref{3}) for
the gluon propagator, bearing in mind that the same results could be
recovered in the limit of small vacuum correlation length  
in the theory with function $D$ instead of ${\tilde D}$ (see equation 
(\ref{5})), 
which is evidently the two dimensional analogue of the ordinary QCD.

Substituting the heavy antiquark Green function $S_{\bar Q} (x,y)$
in the form
\be
S_{\bar Q}(t,x,y)= \left(-i\frac{1+\gamma_0}{2} \theta (-t)
e^{iMt}-i\frac{1-\gamma_0}{2}\theta(t)e^{-iMt}\right)\delta(x-y)
\label{7}
\ee
into (\ref{1}) and performing the integration over $A_{\mu}$ we 
arrive at the following Schwinger--Dyson equation
in the large $N_C$ limit (see {\it e.g.} 
\cite{3} for details):
\be
(i\hat{\partial}_x-m)S(x,y)+\frac{iN_C}{2}\int
d^2z
\gamma_0S(x,z)\gamma_0K(x,z)S(z,y)=\delta^{(2)}(x-y),
\label{8}
\ee
where $S(x,y)=\frac{1}{N_C}S_{\alpha}^{\alpha}(x,y)$ is the colour trace 
of the light quark Green function. Note that as the Green function (\ref{7})
of the heavy antiquark is unity in the colour space in gauge (\ref{2}), 
quantity $S(x,y)$ completely defines the propagation of the colourless
$q\bar Q$ system.

A more simple version of this approach was suggested in \cite{35},
where the full quark Green function in the kernel of equation (\ref{8})
was replaced by the free one, $\gamma_0S\gamma_0\to\gamma_0S_0\gamma_0$.
As we shall see below, such replacement can be justified only for the 
case of heavy quarks. 

In what follows we shall demonstrate that equation (\ref{8}) 
is equivalent to the well--known 't Hooft one
taken in the limit of one heavy mass. The correct value of the chiral
condensate will also be recovered in the given approach. 

The usual way of the 't Hooft equation derivation in any gauge 
consists of two
steps. First the quark self energy part should be calculated to be
substituted into the exact quark propagator. Then a Bethe--Salpeter
equation for the vertex function should be written out, which is in
fact the ultimate answer up to some algebraic transformations. Such
a way of acting hints us that as we managed to combine these both
steps together when deriving equation (\ref{8}), a kind of
separation of the interaction should be present in it. This idea
becomes most clear if one looks more carefully at correlator
(\ref{3}). Indeed, $K$ can be naturally broken into two parts
\be
K=K^{(1)}+K^{(2)},
\label{9}
\ee
where $K^{(1)}(x_0-y_0,x-y)=\frac{g^2}{2}\delta(x_0-y_0)|x-y|$
describes the local part of interaction and
$K^{(2)}(x_0-y_0,x,y)=-\frac{g^2}{2}\delta(x_0-y_0)(|x|+|y|)$ is
nonlocal. 
As we shall work in a more convenient momentum space, we obtain for
$K^{(1)}$ and $K^{(2)}$ correspondingly
\be
K^{(1)}(p_0-q_0,p-q)=-\frac{(2\pi)^2g^2}{p^2}\delta^{(2)}(p-q),
\label{10}
\ee
\be
K^{(2)}(p_0-q_0,p,q)=(2\pi)^2g^2\delta (p_0-q_0)\left(
\frac{\delta(p)}{q^2}+\frac{\delta(q)}{p^2}\right).
\label{11}
\ee

It is easy to see that equation (\ref{8}) with $K^{(1)}$
substituted instead of the full kernel $K$ is nothing but the self energy
equation in the Coulomb gauge $A_1^a=0$ 
\footnote{Note that the self energy is a gauge--variant value.}\cite{2}.

Indeed, with the definition
\be
\begin{array}{l}
S^{(1)}(p_{10},p_1,p_{20},p_2)=(2\pi)^2\delta^{(2)}(p_1-p_2){\tilde S}
(p_{10},p_1),\\
{}\\
{\tilde S}^{-1}(p_0,p)=\gamma_0p_0-\gamma_1p-m-\Sigma(p),
\end{array}
\label{100}
\ee
and the parametrization of the polarization operator $\Sigma(p)$ as
\be
\Sigma(p)=\left[E(p)cos\theta(p)-m\right]+\gamma_1\left[E(p)sin\theta(p)-p
\right]
\ee
equation (\ref{8}) can be reduced to the system of two coupled equations
\cite{2}
\be
E(p)cos\theta(p)=m+\frac{f}{2}\vpint\frac{dk}{(p-k)^2}
cos\theta(k)
\label{15}
\ee
\be
E(p)sin\theta(p)=p+\frac{f}{2}\vpint\frac{dk}{(p-k)^2}
sin\theta(k),
\label{16}
\ee
where 
$$
f=\frac{g^2N_C}{4\pi}=\frac{\sigma}{\pi}.
$$

Solutions of system (\ref{15}), (\ref{16}) are known \cite{2,7}, 
and we shall return to them to calculate the chiral condensate.
Note that in the large mass limit these equations can be solved 
perturbatively, defining in the leading order the dispersion law
of the free quark, $E(p)=\sqrt{p^2+m^2},\;cos\theta(p)=\frac{m}
{E(p)}$, so that Green function $S^{(1)}$ becomes the
free Green function. Such a procedure is obviously legitimate only for
the heavy quark.

Unfortunately, the full equation (\ref{8}) is nonlinear, so any
separation in it is not straightforward.
Our present goal is to justify the separate consideration of 
$K^{(1)}$ we have carried out, and to obtain an equation on the
eigenenergies which is induced by $K^{(2)}$.

The Green function (\ref{100}) together 
with (\ref{10}), (\ref{11}) gives rise to the set of 
propagators presented at Fig.1, where all $K^{(1)}$ loops are already 
absorbed by quark propagator (\ref{100}). 

Let us consider the class of rainbow diagrams shown at Fig.2(a) 
with an arbitrary number of $K^{(2)}$ inner loops. 
Simple but tedious calculations show that all such diagrams vanish (see 
Appendix for the details). The same conclusion holds true for the diagrams
with $K^{(1)}$ covering line (see Fig.2(b)), as the only role played by 
the covering line is providing the integration over $k_0$.
The latter observation means that 
the contribution of the term $K^{(2)}$ is described by the only diagram
depicted at Fig.3, so that equation (\ref{8}) takes the form
$$
(\hat{q}_1-m-\Sigma(q_1))S(q_1,q_2)+\hspace*{8cm}
$$
\be
+\frac{ig^2N_C}{2}\int
\frac{d^2p}{(2\pi)^2}\frac{d^2k}{(2\pi)^2}
\gamma_0 S^{(1)}(p)\gamma_0S(p+k,q_2)K^{(2)}(q_1-p_1,k)=(2\pi)^2\delta^{(2)}
(q_1-q_2),
\ee 
that can be rewritten as 
the Dirac equation for the light quark wave function $\varphi(p_0,p)$.
$$
p_0\varphi(p_0,p)=\gamma_5E(p)sin\theta(p)+\gamma_0E(p)cos\theta(p)-
$$
\be
-\frac{f}{2}\gamma_0\vpint dk[cos\theta(k)+\gamma_1sin\theta(k)]
\frac{\varphi(p_0,k)}{(p-k)^2}-
\label{195}
\ee
$$
-\frac{f}{2}\gamma_0\vpint dk[cos\theta(p)+\gamma_1sin\theta(p)]
\frac{\varphi(p_0,k)}{(p-k)^2}.
$$

After the Foldy--Wouthoysen transformation
\be
{\tilde \varphi}(p_0,p)=T(p)\varphi(p_0,p),\quad T(p)=e^{\frac12\theta(p)
\gamma_1}
\label{200}
\ee
we arrive at the Schr{\"o}dinger--type equation
\be
\varepsilon_n\varphi^0_n(p)=E(p)\varphi^0_n(p)-
f\vpint\frac{dk}{(p-k)^2}cos\frac{\theta(p)-\theta(k)}{2}\varphi^0_n(k),
\label{21}
\ee
which defines the spectrum of the bound states of light quark and heavy 
antiquark, and wave functions of the positive- and negative-energy states
are given by 
\be
{\tilde \varphi}_n^{(+)}(p)=\varphi_n^0(p){1\choose 0},\quad
{\tilde \varphi}_n^{(-)}(p)=\varphi_n^0(p){0\choose 1}.
\ee

Equation (\ref{21}) coincides with the one-body limit of 
the equation obtained in \cite{2}, which in turn is proved to be 
equivalent to the 't Hooft one on the grounds of the Poincar{\'e} 
invariance \cite{2}, numerically \cite{7} and in the framework of 
discrete light-cone quantization \cite{8}. 

Now, when the formal correctness of equation (\ref{8}) is proved, we are ready 
to perform a next step and to show how the chiral condensate appears in quite 
a natural way in the given technique. 

The chiral condensate can be calculated as
\be
<{\bar q}q>=-i\trr S(x,y),
\ee
where the trace is taken both over colour and spinor indices.

At first glance it has nothing to do with the presence of
the heavy antiquark source, but due to the peculiarities of the modified 
Fock--Schwinger gauge we can reformulate our 
spectroscopic problem (\ref{8}) as a problem for the Green function of the 
light quark in gauge (\ref{2}). The colour trace of this Green function
can be constructed from the solutions of equation (\ref{21}) as
$$
N_CS_{ik}(x_0-y_0,x,y)=-iN_C\sum_n\psi_{ni}^{(+)}(x)\bar{\psi}_{nk}^{(+)}(y)
e^{-i\varepsilon_n(x_0-y_0)}\theta(x_0-y_0)+
$$
\be
+iN_C\sum_n\psi_{ni}^{(-)}(x)\bar{\psi}_{nk}^{(-)}(y)
e^{i\varepsilon_n(x_0-y_0)}\theta(y_0-x_0),
\label{101}
\ee
where $i$ and $k$ are spinor indices and
\be
\psi_{ni}^{(\pm)}(x)=\frac{1}{\sqrt{2\pi}}\int_{-\infty}^{\infty} 
dp\;\varphi_{ni}^{(\pm)}(p)
e^{ipx},
\ee
\be
\begin{array}{l}
\varphi_n^{(+)}(p)=T^+(p){\tilde \varphi}_n^{(+)}(p)=
\varphi_n^0(p)T^+(p){1\choose 0},\\
{}\\
\varphi_n^{(-)}(p)=T^+(p){\tilde \varphi}_n^{(-)}(p)=
\varphi_n^0(p)
T^+(p){0\choose 1}.
\end{array}
\label{102}
\ee

Substituting solutions (\ref{102}) into (\ref{101}) and making use of the completeness
condition for the set $\{\varphi_n^0(p)\}$,
\be
\sum_n\varphi^0_n(p)\varphi^{0*}_n(p')=\delta(p-p'),
\ee
we straightforwardly calculate the chiral condensate to be
\be
<\bar q q>=-\frac{N_C}{\pi}\int_0^\infty dp\; cos\theta(p).
\label{103}
\ee

Such way the question whether the chiral symmetry is broken or not depends
on the solution for $\theta(p)$ of system (\ref{15}), (\ref{16}). Solution
found in \cite{2} for the case of $m=0$ reads
\be
\theta(p)=\left\{ 
\begin{array}{lll}
\pi/2&{\rm for}& p>0\\
-\pi/2&{\rm for}& p<0 ,
\end{array}
\right.
\label{29}
\ee
yielding zero value of the condensate. 

In the meantime it is easy to see that there is another solution of system
(\ref{15}), (\ref{16}) which has the following asymptotical forms
\be
\theta(p)\to\left\{
\begin{array}{ll}
\frac{\ds \sqrt{m^2+4\sigma}-m}{\ds 2\sigma}p,&p\to 0\\
{}\\
\frac{\ds \pi}{\ds 2}sign(p),&|p|\to \infty
\end{array}
\right.
\label{30}
\ee

This solution was investigated 
numerically in \cite{7,9}, and in the limit $m\to 0$ 
it gives for the condensate
\be
<\bar{q}q>_{m=0}=-0.29N_C\sqrt{2f}.
\label{31}
\ee

Again we refer here to the discrete light-cone quantization treatment 
\cite{8}, where it was shown that expression (\ref{103}) as well as the
numerical estimate (\ref{31}) coincide with the result obtained in
\cite{10} via operator expansion.

We would like to conclude with a brief comment on the quantum mechanical
aspects of equations (\ref{195}), (\ref{21}). 
With these equations we resolve the long-standing problem of the the Lorentz
structure of the effective confining force. The quasiclassical behaviour
of the spectrum given by equation (\ref{21}) reproduces a Regge trajectory 
with a slope specific for the time--like
vector confinement, $M^2=\pi\sigma n$, but
it is known that for the Dirac particle in the time--like 
external vector field the 
ill--starred Klein paradox takes place. Still there is no puzzle in such a
situation: the interaction in the underlying Dirac equation (\ref{195})
does not contain the time--like vector part at all, there are only scalar
and space--like vector pieces which enter in a certain combination in  
the essentially nonlocal way. We are able to perform the Foldy--Wouthoysen
transformation (\ref{200}) in a closed form, and this fact indicates 
that there is no dangerous light quark Zitterbewegung and therefore there are 
no reasons to be afraid of the Klein paradox.
\bigskip

We acknowledge useful discussions with A.A.Abrikosov Jr., 
K.G. Bo\-res\-kov and Yu.A. Si\-mo\-nov.
\smallskip

This work is supported by grants 96-02-19184a and 97-02-16404 of Russian
Fundamental Research Foundation and by INTAS 94-2851 and 93-0079ext.
\bigskip

{\parindent=0cm\Large\bf Appendix}
\medskip

Anticipating zero result for all diagrams shown at Fig.2 let us omit 
all inessential multipliers keeping only the
terms 
containing $\gamma$-matricies and $k_0$ in order to perform the 
integration over $k_0$ explicitly.
It is easy to see that the general form of the given integrals is
$$
I_n\sim\int_{-\infty}^{+\infty}dk_0
\gamma_0S^{(1)}(k_0,p_1)M(p_1,p_2)S^{(1)}(k_0,p_2)\ldots S^{(1)}(k_0,p_{n-1})
M(p_{n-1},p_n)S^{(1)}(k_0,p_n)\gamma_0,
$$
where $p_1\equiv q$, $p_N\equiv p$
and $M(p,q)$ is defined by the diagram from Fig.3.

Quark propagator (\ref{100}) can be written in the form
$$
S^{(1)}(p_0,p)=\frac{1}{p_0\gamma_0-\Omega(p)},\quad
\Omega(p)=-E(p)e^{-\theta(p)\gamma_1},
$$
where $\Omega(p)$ possesses the following properties
$$
\Omega^+(p)\Omega(p)=E^2(p), \quad \gamma_0\Omega(p)=\Omega^+(p)
\gamma_0.
$$

Using these properties of the function $\Omega(p)$ and omitting all 
odd powers of $k_0$ in the nominator, 
which obviously give no contribution to $I_n$, we 
arrive at the following expression for $I_n$:
$$
I_n\sim\int_{-\infty}^{+\infty}dk_0 \frac{k_0^n+k_0^{n-2}
\sum_{i\ne j}^n E_iE_j+\ldots+\prod_{i=1}^n E_i}{
(k_0^2-(E_1-i\varepsilon)^2)(k_0^2-(E_2-i\varepsilon)^2)\ldots
(k_0^2-(E_n-i\varepsilon)^2)},
$$
for even $n$, or\hspace*{12.8cm}$(B.2)$
$$
I_n\sim\int_{-\infty}^{+\infty}dk_0 \frac{k_0^{n-1}\sum_{i=1}^nE_i+k_0^{n-3}
\sum_{i\ne j}^n E_iE_j+\ldots+\prod_{i=1}^n E_i}{
(k_0^2-(E_1-i\varepsilon)^2)(k_0^2-(E_2-i\varepsilon)^2)\ldots
(k_0^2-(E_n-i\varepsilon)^2)},
$$
for odd $n$, where $E_i\equiv E(p_i)$.

As odd powers of $k_0$ give no contribution to the integral, we can add
the appropriate terms to the numerator in order to come either to
$\prod_{i=1}^n(k_0+E_i)$ or to $\prod_{i=1}^n(k_0-E_i)$. This means that only
one of two series of poles survives, either in the upper, or in the lower 
half-plane. 
Closing the contour of integration in the complex plane via the half-plane 
without poles we arrive at the zero result for $I_n$ for any $n$.

\begin{center}
\begin{figure}[t]
\epsfxsize=17cm
\epsfbox{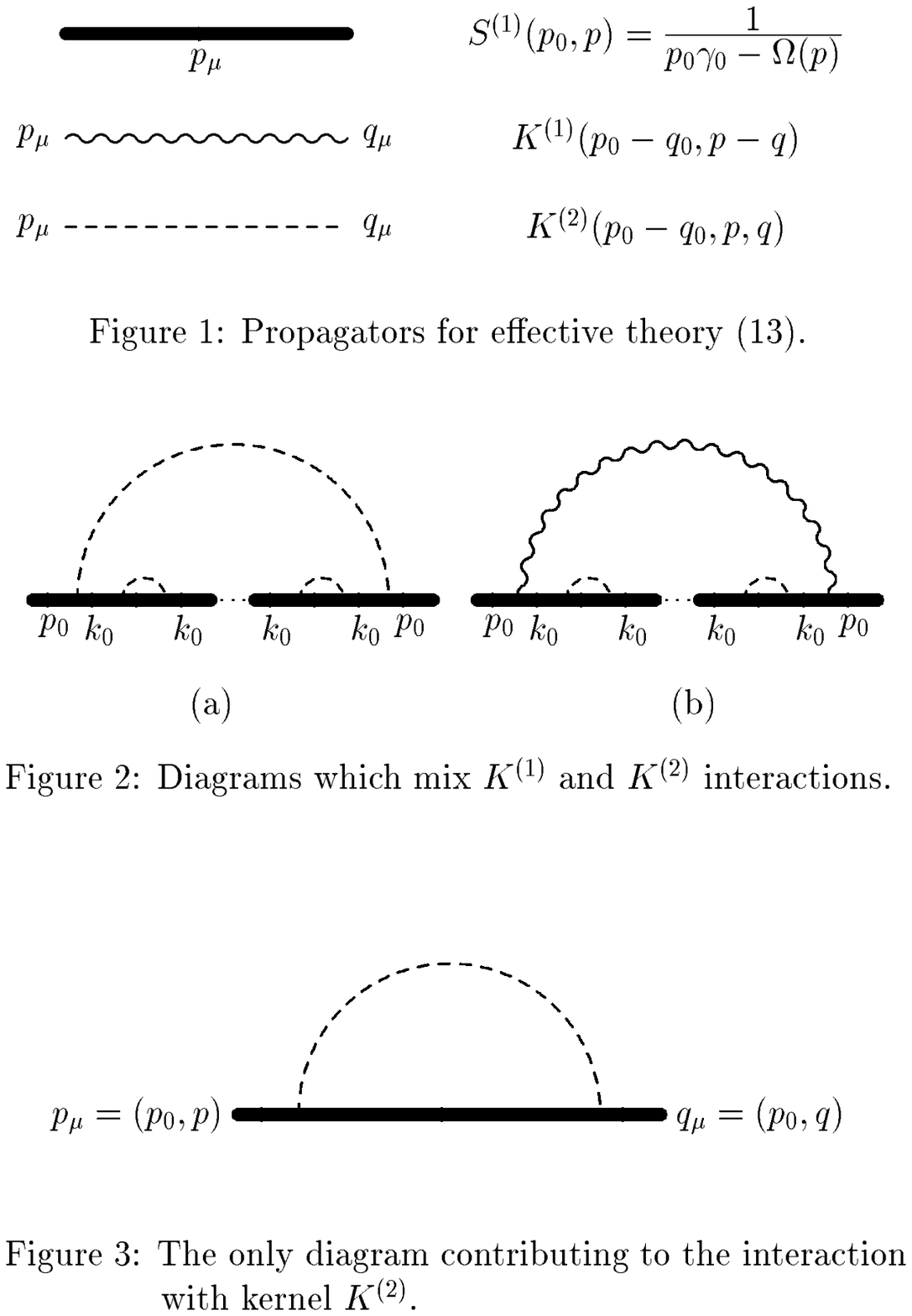}
\end{figure}
\end{center}
\end{document}